# Above-ground biomass change estimation using national forest inventory data with Sentinel-2 and Landsat 8

Puliti, S. [a] *, Breidenbach, J. [a], Schumacher, J. [a], Hauglin, M. [a], Klingenberg, T.F. [b], Astrup, R. [a]

[a] Norwegian Institute for Bioeconomy Research (NIBIO); Division of Forest and Forest Resources; National Forest Inventory. Høgskoleveien 8, 1433 Ås, Norway;

[b] Norwegian Mapping Authority (Kartverket), Land Mapping Division, P.O. Box 600, Sentrum, 3507, Hønefoss, Norway

* Correspondence: stefano.puliti@nibio.no;

**Abstract**

This study aimed at estimating total forest above-ground net change (ΔAGB; Mt) over five years (2014 – 2019) based on model-assisted estimation utilizing freely available satellite imagery. The study was conducted for a boreal forest area (approx. 1.4 Mill hectares) in Norway where bi-temporal national forest inventory (NFI), Sentinel-2, and Landsat data were available. Biomass change was modelled based on a direct approach. The precision of estimates using only the NFI data in a basic expansion estimator were compared to four different alternative model-assisted estimates using 1) Sentinel-2 or Landsat data, and 2) using bi- or uni-temporal remotely sensed data.

We found that the use of remotely sensed data improved the precision of the purely field-based estimates by a factor of up to three. The most precise estimates were found for the model-assisted estimation using bi-temporal Sentinel-2 (standard error; SE= 1.7 Mt). However, the decrease in precision when using Landsat data was small (SE= 1.92 Mt). In addition, we found that ΔAGB could be precisely estimated also when remotely sensed data were available only at the end of the monitoring period.

We conclude that satellite optical data can considerably improve ΔAGB estimates, even in those cases where repeated and coincident NFI data are available. The free availability, global coverage, frequent update, and long-term time horizon make data from programs such as Sentinel-2 and Landsat a valuable data source for a consistent and durable monitoring of forest carbon dynamics.

# 1. Introduction

Forests play a central role in regulating the global climate through processes such as carbon uptake, carbon emission, and the regulation of the water and energy cycles (Herold et al. 2019). Forest structures are dynamic systems varying through space and time. Understanding the dynamics of the forest above-ground biomass (AGB) is critical to comprehend better the magnitude of the impact that forest dynamics and forest management have on climate change (Eggleston et al. 2006). Repeated national forest inventory (NFI) data, along with freely available medium-resolution satellite imagery such as Sentinel-2 (Drusch et al. 2012) or Landsat data (Wulder et al. 2012), offer unique possibilities for long-term monitoring of AGB dynamics (GFOI 2020). Such medium-resolution satellite data are nowadays available at a global scale, at short intervals in time (5-16 days at the equator), and their acquisition is planned to span across several decades, making them one of the most useful source of auxiliary information for national and international forest monitoring programs. Despite the current need for globally consistent estimates of forest AGB dynamics (Duncanson et al. 2019), little is known about the contribution of satellite optical data to estimate AGB changes in green-house gases inventories.

## 1.1 Remotely sensed based ΔAGB estimation

During the past decade, the majority of the studies aimed at developing ways to map and estimate AGB stocks at specific points in time (Zolkos et al. 2013), and only a minority of studies looked at the use of multi-temporal remotely sensed data for estimating and mapping AGB change (ΔAGB). As a prominent source of auxiliary information for AGB estimation and mapping, airborne laser scanning (ALS) data has been found useful also in improving the precision of ΔAGB estimates based on field reference data alone by factors of 2 – 9 (Næsset et al. 2013; Skowronski et al. 2014; McRoberts et al. 2015; Næsset et al. 2015; Bollandsås et al. 2018). Other three-dimensional remotely sensed data such as digital aerial photogrammetry, have been found useful to fit nationwide ΔAGB models (Price et al. 2020). Even though such three-dimensional remotely sensed data represent the state-of-the-art, their use remains limited in terms of area coverage due to the large costs and the logistical complexity intrinsic of aerial data acquisition campaigns.

When aiming at producing consistent estimates on a continental or even global scale, satellite data represent a more cost-efficient and thus promising data source for the estimation of ΔAGB over extended periods (GFOI 2020). New and upcoming scientific satellite missions such as the GEDI and ESA's BIOMASS have been specifically tailored to the need for improving our understanding of near-global forest biomass stocks (Duncanson et al. 2019). Even though such missions may improve our understanding of the present state of forest AGB stocks, they remain characterized by short lifespans (2 – 5 years) and do not cover the entire globe. Amongst the past experiences, both interferometric synthetic aperture radar (InSAR) (Solberg et al. 2014; Karila et al. 2019) and optical data (Powell et al. 2010; Main-Knorn et al. 2013), such as TanDEM-X and Landsat multi-temporal data have been used to map ΔAGB. Even though InSAR data has the advantages of being cloud-insensitive and characterizing the forest canopy height, due to the commercial nature of available InSAR data (e.g. TanDEM-X) their use remains limited across space and time.

Amongst the broad panorama of the currently available remotely sensed data, medium-resolution (10 – 30 m) optical data represents one of the most promising source of auxiliary data as they allow virtually anyone to access near-real-time as well as archive imagery over any point on earth and free of cost. The long term lifespan of these missions makes them particularly suitable for continuous monitoring of forest AGB dynamics.

*1.2 ΔAGB using open satellite optical data*

In the realm of optical data, time series of reflectance data from Landsat have been and continue to be effectively used to monitor the global extent and patterns of forest cover disturbance and recovery (Cohen et al. 2010; Hansen et al. 2013; White et al. 2018). Concerning the quantification of the carbon uptake and emission, methods to estimate AGB losses from carbon density maps have been proposed (Baccini et al. 2012; Csillik et al. 2019). While such maps may be used to quantify forest carbon emissions, they lack information on the carbon sequestration potential and thus the net ΔAGB. Only a handful of studies used time-series of Landsat data to model AGB over time and consequently derive AGB changes (Powell et al. 2010; Main-Knorn et al. 2013; Matasci et al. 2018; Nguyen et al. 2020; Wulder et al. 2020). Even though useful to characterize areal changes based on the loss or accumulation of forest AGB, the previously proposed methods lacked a rigorous assessment of the uncertainty of the ΔAGB estimates and thus do not comply with guidelines for estimation of emissions and removals of greenhouse gasses in forests indicated by GFOI (2020). Knowledge of the uncertainty of the ΔAGB estimates is essential for current green-house gasses inventories, and according to good practice, the estimates should be neither systematically over- nor under-estimating the true ΔAGB (IPCC 2019).

*1.3 Methods for change estimation*

In this study, we assessed methods to attain ΔAGB estimates based on the IPCC recommendations (Eggleston et al. 2006) and the GFOI methods and guidance (GFOI 2020). In forest inventories supported by remotely sensed data, the estimation of ΔAGB and its uncertainty may be done either through model-assisted or model-based estimators (Gregoire 1998; Ståhl et al. 2016). Both rely on the use of models linking field reference data with remotely sensed auxiliary data to improve the precision of purely field-based estimates. The model-assisted estimator is preferred in those cases where a probability sample of field observations is available as it is nearly-unbiased even if the model has a lack of fit (Särndal 1984).

Concerning the models used, ΔAGB may be estimated either using a direct or indirect modeling approach (McRoberts et al. 2015). The former consists in directly modeling ΔAGB, while the latter consists in modeling AGB state in the two points in time ($T_1$ and $T_2$) separately and estimating ΔAGB as the difference between the AGB in $T_2$ and the AGB in $T_1$. When repeated and coincident field plot data are available, the direct approach is often preferred as it relies on a single model and thus a single error source (Fuller et al. 2003; Bollandsås et al. 2013; Skowronski et al. 2014; McRoberts et al. 2015). Within the direct approach, one may consider using either bi-temporal or only uni-temporal remotely sensed data. The use of the bi-temporal approach represents a familiar approach (McRoberts et al. 2015; Næsset et al. 2015; Bollandsås et al. 2018), which intuitively relies on temporally matching field and remotely sensed data in both $T_1$ and $T_2$. The uni-temporal approach is representative of those cases when remotely sensed data are available only for $T_2$. Thus, the changes in AGB are explained only as a function of a snapshot (e.g., satellite image) at the end of the monitoring period. While unexplored, this approach is relevant for satellite optical data in those cases without available images in $T_1$ due to cloud-cover or to the fact that the satellite was not even launched. Direct ΔAGB models relying on uni-temporal remotely sensed data have the potential to broaden the possibility to estimate ΔAGB over a wider time window.

*1.4 Field reference data*

An important limiting factor to direct ΔAGB modeling lies in the lack of repeated and coincident field reference data at different points in time (Næsset et al. 2015). Consequently, often the gain-loss method represents the only solution (GFOI 2013; McRoberts et al. 2020). Continuously updated NFIs represent a unique data source to attain Tier 3 information based on a stock-difference method. As an example, the Norwegian NFI is composed of a systematic network of more than 22,000 permanent field plots. Every five years, a fifth of the plots are re-measured, allowing to estimate ΔAGB over the past five years period (Breidenbach et al. 2020a). While the scope of NFIs is mainly to provide nationwide and regional statistics, the availability of repeated and geolocated AGB observations over entire nations offers unique possibilities for the direct estimation of ΔAGB using satellite optical data and thus can offer an important contribution to global efforts to monitor forest carbon dynamics.

*1.5 Study objective*
The objective of this study was to estimate ΔAGB for a period of five years (2014 – 2019) using bi-temporal NFI data and either bi-temporal ($T_1$ = 2015 and $T_2$ =2019) or uni-temporal ($T_2$ =2019) Sentinel-2 or Landsat data. Direct estimates of the total net ΔAGB were obtained according to a model-assisted estimator and the results compared to estimates based on a basic expansion estimator (BE) using the NFI data alone.

**2. Materials and methods**

*2.1 Study area*
The study area was in south-eastern Norway and comprised a total area of 13,659 km² (see Figure 1). This specific area was selected as it was one of the largest contiguous area in Norway where Sentinel-2 data were available and free of cloud cover during 2015, which was also the first year that the Sentinel-2 A satellite was operational (i.e. limited number of images). The forest area estimated using all NFI plots within the AOI is nearly 70% of the total area. The AOI is characterized by a large proportion of forest (nearly 70%) that are actively managed for timber production. With the absence of substantial forest fires or insect outbreaks in the study area between 2014 and 2019, forest harvest represents the primary source of disturbance causing the reduction in AGB stocks.

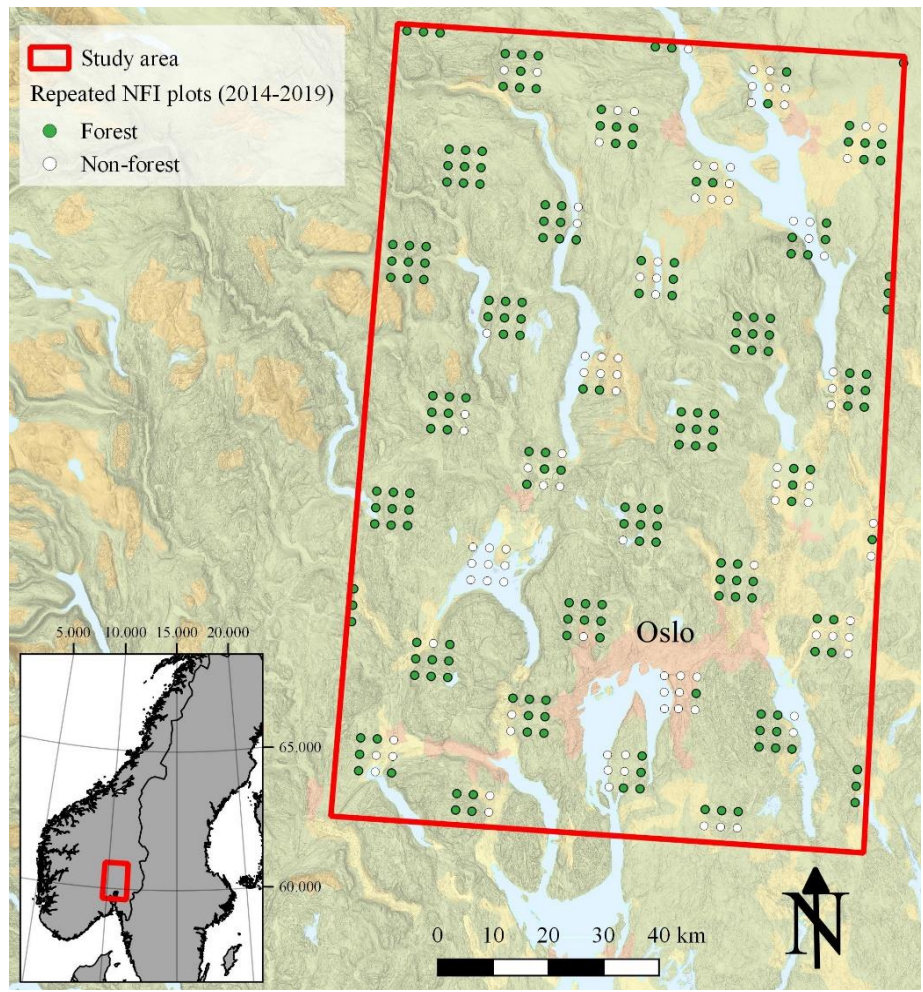

**Figure 1**. Overview of the study area, highlighting the NFI sampling design for the permanent plots measured in 2014 and 2019.

*2.2 National Forest Inventory data*

The Norwegian NFI is based on a five-year repeat cycles, according to which, a fifth of the plots are revisited every fifth year. This allows to calculate the plot-wise ΔAGB as the difference between the AGB stock at the end ($T_2$) and at the beginning of the monitoring period ($T_1$). Permanent NFI field plots are systematically located in a 3×3km grid within the study area. The set of field plots measured each year is selected based on a latin-square design, resulting in evenly distributed clusters of 3×3 field plots measured each year (Fig. 1). The permanent field plots measured in 2014, and again in 2019 were used in the present study.

2.2.1 AGB estimation on the NFI field plots

Within the 250 m$^2$ circular field plots, diameter at breast height (DBH) of trees with DBH >= 5 cm was recorded. Tree height was measured on selected trees, according to the NFI field protocol (Breidenbach et al. 2020a). Field measurements from 2014 and 2019 were then used with species-specific allometric models to estimate AGB for each field plot at $T_1$ and $T_2$, respectively (Marklund 1988). The center point of each field plot was positioned using a Global Navigation Satellite System receiver, allowing for spatial alignment with satellite data.

Further details about the field registrations in the Norwegian NFI can be found in (Breidenbach et al. 2020a)

*2.3 Remotely sensed data*

2.3.1 Sentinel-2

Atmospherically corrected Sentinel-2 (level 2A) cloud-free mosaics were generated for two points in time: August 2015 and August 2019. While optimally, the remotely sensed data should be temporally matching the field reference data in 2014 and 2019, the first Sentinel-2 images were available only in 2015. The mosaicking method is described in detail in the study by Puliti et al. (2020). For each of the NFI plots, we extracted the sentinel-2 band values corresponding with the coordinates of the plot centre.

2.3.2 Landsat

Landsat Analysis Ready Data produced by the Global Land Analysis and Discovery team at the University of Maryland (GLAD ARD, Potapov et al. 2020) were used as a source of yearly cloud-free Landsat mosaics. For comparison with the Sentinel-2 data, we selected mosaics from the same two points in time (i.e. 2015 and 2019). These data represent a globally available data source for long term monitoring (1997-present) of land cover change mapping. An advantage of using these data lies in the fact that GLAD ARD data are produced consistently at a global scale and thus are particularly suitable to ensure transparency and reproducibility of the estimates.

The cloud-free mosaics of normalized surface reflectance from Landsat data were generated using the phenological metrics type A in the GLAD tools for Landsat ARD applications (GLAD 2020). The mosaics are gap-filled to remove clouds, snow, and shadows. Each composite represents the average of reflectance values between the $25^{th}$ and $75^{th}$ percentile from the gap-filled cloud-free annual observation time-series.

Specific details of the processing steps involved in the production of GLAD ARD data can be found in Potapov et al. (2020). As for the Sentinel-2 data, the bands' values corresponding to each NFI plot centre were extracted.

2.3.3 Forest Mask

A nationally available forest mask was used to geographically define the population of interest. The forest mask is based on the Norwegian land capability classification system, "AR5" (Ahlstrøm et al. 2014) and is currently used for the nationwide mapping of forest resources in Norway (Astrup et al. 2019). While dating back to the 1960s, such a forest mask currently represents the best source of information on the extent and spatial distribution of forests in Norway with an accuracy of 92% (Breidenbach et al. 2020b). The forest mask is produced from a combination of field and aerial surveys and is currently being updated based on ALS data.

*2.4 Statistical analysis*

2.4.1 Modelling ΔAGB

In order to model ΔAGB, we opted for a direct method according to which the remotely sensed auxiliary variables are used to model the change in AGB in the NFI plots directly. Two different approaches to modeling ΔAGB, namely:

1. **Bi-temporal**: such an approach consisted of using explanatory variables from Sentinel-2 and Landsat data from both points in time (2015 and 2019).
2. **Uni-temporal**: this approach assumed that remotely sensed data were available only for $T_2$ and thus links the changes in AGB to the present state (2019) as pictured by satellite optical data.

The models were fitted using explanatory variables independent to pairing, meaning that the bands selected for $T_1$ are not necessarily selected for $T_2$ and vice versa. This model form was adopted as it was found to be the one yielding the most precise estimates in a comparative study of several different model forms by McRoberts et al. (2015). Using only the plots classified as forest in the NFI data, we fitted four separate multiple linear models linking the ΔAGB from the NFI data with bi-temporal or uni-temporal data either from Sentinel-2 or Landsat. The explanatory variables consisted of the bands' values as well as band indices. Similarly to the normalized difference vegetation index (NDVI), the band indices were calculated as the ratio between the difference and the sum of band pairs. We calculated the band indices from all band combinations (i.e., 45 and 15 combination for Sentinel-2 and Landsat for each point in time).

**Table 1**. Spectral bands (nm) and resolution (m) for the Sentinel-2 and Landsat imagery.

| Band number[a] | Sentinel-2 | Landsat |
|---|---|---|
| B2 | Blue (458 – 523 nm) 10 m | Blue (452 – 512 nm) 30 m |
| B3 | Green (543 – 578 nm) 10 m | Green (533 – 590 nm) 30 m |
| B4 | Red (650 – 680 nm) 10 m | Red (636 – 673 nm) 30 m |
| B5 | Red-edge 1 (698 – 713 nm) 20 m | NIR (851 – 879 nm) 30 m |
| B6 | Red-edge 2 (733 – 748 nm) 20 m | SWIR 1 (1566 – 1651 nm) 30 m |
| B7 | Red-edge 3 (773 – 793 nm) 20 m | SWIR 2 (2107 – 2294 nm) 30 m |
| B8 | NIR 1 (785 – 899 nm) 10 m | - |
| B8A | NIR 2 (855 – 875 nm) 20 m | - |
| B11 | SWIR 1 (1565 – 1655 nm) 20 m | - |
| B12 | SWIR 2 (2100 – 2280 nm) 20 m | - |

[a] the band numbers reported for Sentinel-2 and Landsat data correspond to the original order for the respective sensors as described in their respective official documentation

The predictor variables were selected with attention to ensure parsimonious models and avoid multicollinearity. First, we performed a branch-and-bound search for the best subset based on the Bayesian information criterion. Second, the models were penalized for multicollinearity by ensuring that the largest variance inflation factor was less than five.

2.4.2 Estimation

An estimate of the total forest net ΔAGB ($\hat{t}$) and its variance ($\widehat{Var}(\hat{t})$) using a basic expansion (BE) estimator just based on field data and a model-assisted (MA) estimator were calculated as presented in the following.

We assumed the field reference data ($S$) to be a simple random sample of size $n$, resulting in the BE estimator of the total

$$\hat{t}_{BE} = \frac{A}{n} \sum_{i \in S=1}^{n} y_i \quad (1)$$

where $A$ =1,342,397 ha is the total land area of the study area, and $y_i$ is the ΔAGB over a five years period (t ha$^{-1}$), for the $i$th NFI field plot ($i$ = 1, …, $n$). The variance of the total was estimated by

$$\widehat{Var}(\hat{t}_{BE}) = \frac{S^2}{n} A^2 \quad (2)$$

where $S^2$ is the sample variance

$$S^2 = \frac{1}{n-1} \sum_{i=1}^{n} (y_i - \bar{y})^2 \tag{3}$$

and $\bar{y}$ is the sample mean.

For MA estimation, the ΔAGB models devised in sub-section 2.4.1 using bi- or uni-temproal data from Sentinel-2 or Landsat data were applied to each pixel in the study area. The pixel-predictions were then masked using the available forest mask to exclude non-forest areas resulting in a synthetic estimate of ΔAGB. Furthermore, a binary indicator variable $I$ defined whether the plot was forested ($I$=1) or not ($I$=0) according to the NFI. The MA estimator of total ΔAGB was

$$\hat{\tau}_{MA} = A \frac{1}{N} \sum_{k=1}^{N} \hat{y}_k + A \frac{1}{n} \sum_{i=1}^{n} \varepsilon_i \tag{4}$$

where $\hat{y}_k$ are pixel-level ΔAGB predictions, N is the number of pixels within the forest mask and $\varepsilon_i$ is the residual

$$\varepsilon_i = y_i - (\hat{y}_i I_i) \tag{5}$$

where $\hat{y}_i$ is the plots' predicted ΔAGB and $I_i$ the indicator variable. The first component of eq. 4 represents the synthetic (map-based) estimator and the second component is an estimated correction factor. The MA variance was estimated according to

$$\widehat{Var}(\hat{\tau}_{MA}) = \frac{A^2}{n(n-1)} \sum_{i=1}^{n} (\varepsilon_i - \bar{\varepsilon})^2 \tag{6}$$

where $\bar{\varepsilon}$ is the mean residual.

Finally, the BE and MA estimates were compared by the relative efficiency (RE), calculated as the ratio $\widehat{Var}(\hat{\tau}_{BE})/\widehat{Var}(\hat{\tau}_{MA})$ where the latter was estimated either using Sentinel-2 or Landsat data. The RE was used to describe the relative improvement in the precision of the model-assisted estimates over the direct estimates. As the population of study included only forest areas, the ΔAGB for plots outside forest areas (according to the NFI classification) was set to zero (Breidenbach et al. 2020b).

**3 Results and discussion**

*3.1 Models*

The bi-temporal Sentinel-2 model (see sub-section 2.4.1) included the index between the red-edge 3 (B7) and SWIR 2 (B12) in 2015 and 2019 (Table 2). The selection of the same pair of bands in bi-temporal Sentinel-2 data is intuitive as it allows to link changes in AGB to changes in reflectance. On the other hand, the selected variables in the Landsat bi-temporal model were not paired and included the blue (B2) from 2015, and the red (B4) and the NIR band (B7) for 2019. The Sentinel-2 and Landsat variables of $T_2$ were considerably stronger correlated to ΔAGB than those in $T_1$ justifying the adoption of a uni-temporal approach. The maximum correlations were r=-0.65 for Sentinel-2 SWIR2 and r=-0.6 for Landsat's SWIR2 in $T_2$ and r=-0.15 for Sentinel-2 red and r=-0.06 for Landsat's SWIR2 band in $T_1$. When using only satellite optical data from $T_2$, the Sentinel-2 model included the ratio between the red (B4) and the red-edge 3 (B7), and the ratio between the red and a SWIR1 band (B11). The Landsat uni-temporal model included the NIR (B5) and a SWIR 2 (B7).

**Table 2.** Summary of the ΔAGB models for Sentinel-2 and Landsat

| Auxiliary data | Model type | model | Adj.$R^2$ |
|---|---|---|---|
| Sentinel-2 | Bi-temporal | ΔAGB = $-79.86 - 137.32\ B_{7/12}^{2015} + 284\ B_{7/12}^{2019}$ | 0.64 |
| | Uni-temporal | ΔAGB = $-185.93 - 485.72\ B_{4/7}^{2019} + 301.76\ B_{4/11}^{2019}$ | 0.61 |
| Landsat | Bi-temporal | ΔAGB = $-32.53 + 0.071\ B_2^{2015} - 0.050\ B_7^{2019} - 94.69\ B_{4/7}^{2019}$ | 0.56 |
| | Uni-temporal | ΔAGB = $-0.04 + 0.0095\ B_5^{2019} - 0.04\ B_7^{2019}$ | 0.46 |

The Sentinel-2 models had better model fit (Adj.$R^2$ = 0.61 – 0.64) than the Landsat ones (Adj.R2 = 0.46 – 0.56). The latter was characterized by a greater decrease in the model's explanatory power when using uni-temporal rather than bi-temporal data. Compared to previous studies adopting a direct modeling approach for ΔAGB estimation in boreal forests, the models devised in this study had model fit similar or better than InSAR data ($R^2$ in the range 0.2 – 0.6) (Næsset et al. 2015) or ALS data in montane forests ($R^2$= 0.28) (Bollandsås et al. 2018), but poorer than what was previously found when using ALS data in productive forests (i.e. Adj. $R^2$ in the range 0.6 – 0.9) (Næsset et al. 2013; McRoberts et al. 2015). While ALS data currently are the state-of-the-art source of auxiliary data for forest AGB modelling, its large acquisition costs make it an unsuitable source of information for long-term, continuous, and nationwide AGB monitoring programs.

An analysis of the Sentinel-2 model's predictions (see Figure 2) revealed a tendency of all models to under-predict the losses in plots characterized by a negative ΔAGB according to the NFI data. The Sentinel-2 and the Landsat based ΔAGB predictions were negative for 84% and 69% of the plots with observed AGB loss.

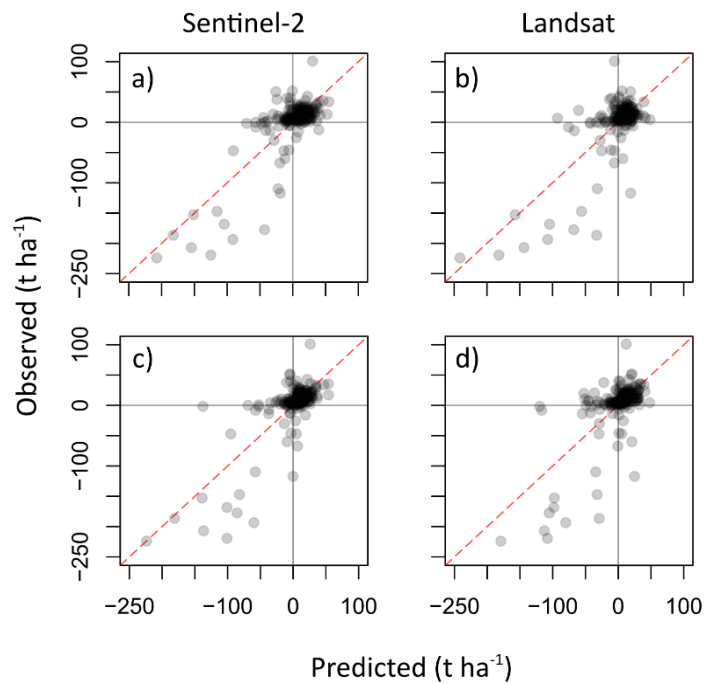

**Figure 2**. Observed (NFI) vs. predicted ΔAGB using either Sentinel-2 (a and c) or Landsat data (b and d) and using bi-temporal data (a and b) or uni-temporal data (c and d). The perpendicular lines represent the 0,0 origin, and the dashed line is the 1:1 line.

Differently from previous studies on direct ΔAGB estimation (Næsset et al. 2013; McRoberts et al. 2015), no truncation of the models' predictions was deemed necessary. Such choice was motivated by the fact that the percentage of pixels in the entire study area where any of the predictor variables were outside the range of values in the sample was limited to a maximum of 0.6% for the Sentinel-2 data and 1.4% for the Landsat data.

*3.2 Estimation*

The BE estimate of ΔAGB based exclusively on NFI data was 0.12 Mt with a SE = 2.94 Mt (Table 3). The total estimate for the studied period was smaller than all ΔAGB from previous five-years NFI cycles from 2010 until 2018 (see Figure 3), mainly due to a larger number of plots with AGB loss during the studied period (2014-2019) compared to the earlier periods. The direct estimate of yearly average ΔAGB was (0.03 t ha$^{-1}$ year$^{-1}$) was smaller than the estimates by Næsset et al. (2013) in productive forests between 2000 and 2010 (1.78 t ha$^{-1}$ year$^{-1}$), the estimates by Bollandsås et al. (2018) in montane forests in the period between 2008 and 2012 (0.21 t ha$^{-1}$ year$^{-1}$) or the estimate by Strîmbu et al. (2017) for an areal extent similar to this study for the period 2006 – 2011 (1.27 t ha$^{-1}$ year$^{-1}$). Even though the abovementioned studies were conducted in different areas and for different periods, the reported estimates are in line with the decreasing ΔAGB trend visible in our study area in the period 2010 – 2020 (Figure 3).

The model-assisted estimates of the total net ΔAGB using the bi-temporal model were of 2.57 Mt (SE = 1.7 Mt) and 1.83 Mt (SE = 1.92 Mt) when using either Sentinel-2 or Landsat, respectively. When using uni-temporal data, the uncertainty of the estimates increased both for the Sentinel-2 (SE = 1.84 Mt) and the Landsat (SE = 2.16 Mt), with the latter being characterized by a steeper decrease. While all the model-assisted estimates for total ΔAGB were larger than the direct estimate, they were well within its 95% confidence interval (see Figure 3) and thus not significantly different from the direct estimate.

**Table 3**. Estimated total change in above-ground biomass ($\hat{t}$; Mg)

| Availability of remotely sensed data | BE | | MA Sentinel-2 | | MA Landsat | |
|---|---|---|---|---|---|---|
| | Tot (Mt) | SE (Mt) | Tot (Mt) | SE (Mt) | Tot (Mt) | SE (Mt) |
| Bi-temporal | 0.12 | 2.94 | 2.57 | 1.7 | 1.83 | 1.92 |
| Uni-temporal | | | 1.56 | 1.84 | 2.33 | 2.16 |

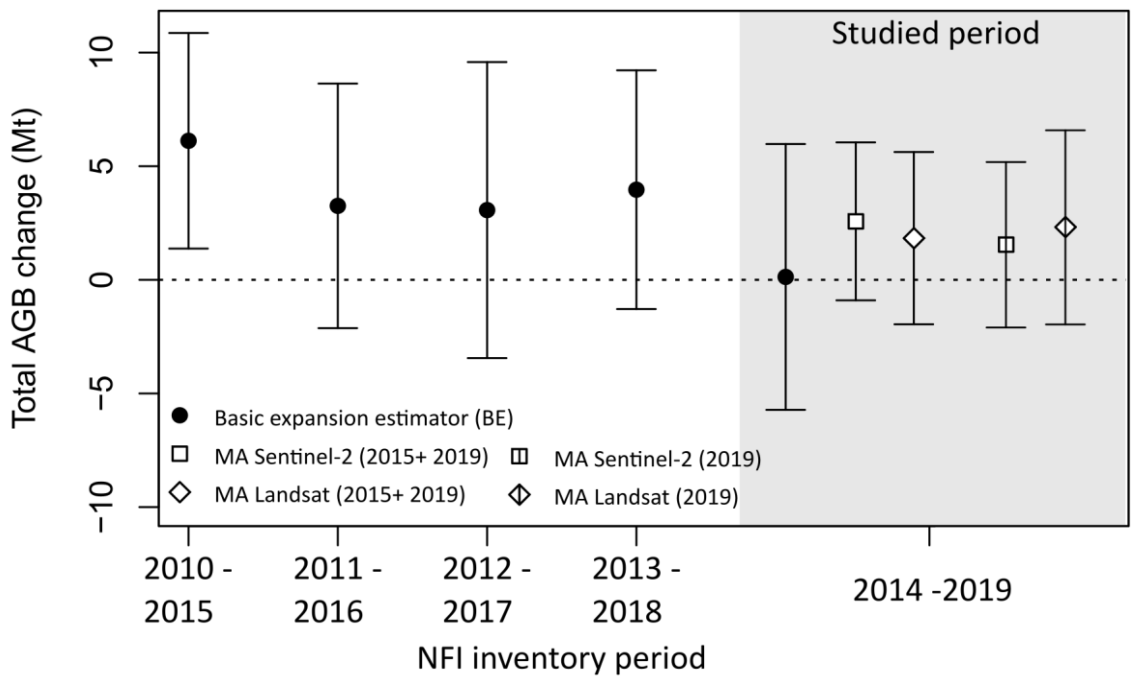

**Figure 3**. Total estimated ΔAGB and 95% confidence intervals for the direct estimate and model-assisted estimates using either Sentinel-2 (MA Sentinel-2) or Landsat data (MA Landsat) and using either bi-temporal or uni-temporal mosaics.

Because the estimators used in this study were either unbiased or nearly unbiased (Särndal 1984), the differences in the point estimates can be attributed to random variations (McRoberts et al. 2015). A potential reason behind such variations may be due to the nature of the satellite cloud-free mosaics, which, by blending multi-date imagery, are characterized by heterogeneous sun-target-sensor geometry and atmospheric conditions. This variation is further exacerbated when using bi-temporal mosaics, due to the compound effect of random variations in reflectance from two points in time.

In this study, Sentinel-2 data for T1 were available only from 2015, and thus the temporal mismatch between the NFI data collection in 2014 and the remotely sensed data may have caused a reduction in the detectable AGB losses. While this temporal mismatch may be a drawback of this study, model-assisted estimates based on Landsat data from 2014 and 2019 (2.38 Mt; SE = 1.89 Mt) were of similar magnitude and without an appreciable increase in precision compared to when using Landsat data from 2015 and 2019. Such a result shows the viability of the proposed method even when the remotely sensed data does not perfectly match the timing of the field reference data acquisition. Such characteristic may be particularly attractive for passive optical data as it allows to draw satellite images from a broader time window, and thus increase the chances of having a pair of cloud-free mosaics.

The relative efficiency of the MA estimates was 2.7 and 2.3 for the estimates using bi-temporal data and either the Sentinel-2 or the Landsat data, respectively (see Table 4). This translates into the need of nearly three times the number of NFI plots to obtain the same level of accuracy as for the model-assisted estimate based on Sentinel-2 data. The relative efficiency found in this study was smaller than the range in REs reported by Næsset et al. (2013), Næsset et al. (2015), and McRoberts et al. (2015) for the direct model-assisted estimation of ΔAGB over a period of 10 years using ALS data (RE= 3.1 – 10). When compared to the RE for InSAR data, our study found a larger RE than an earlier study by Næsset et al. (2015), where the RE was in the range of 1.8 – 2.5.

**Table 4**. Relative efficiency for the model-assisted estimates of ΔAGB using either bitemporal or uni-temporal data from either Sentinel-2 or Landsat.

|              | MA Sentinel | MA Landsat |
| ------------ | ----------- | ---------- |
| Bi-temporal  | 2.7         | 2.3        |
| Uni-temporal | 2.5         | 1.8        |

*3.3 Comparison between Sentinel-2 and Landsat data*

Based on the results of this study, we found that Sentinel-2 data was the best source of auxiliary information. This could be partly attributable to the finer spatial (Mascorro et al. 2015), spectral, and temporal resolution of Sentinel-2 data compared to Landsat data. In addition to data-specific differences, variations due to the mosaicking methods adopted for the two remotely sensed data sources may have affected the model fit of these two models. In particular, the Sentinel-2 mosaic was generated from summer-only acquisitions, while Landsat GLAD ARD mosaic was generated using an average of a gap-filled yearly stack of images, thus introducing a more considerable variation in the reflectance values. By adopting a narrow timeframe, the Sentinel-2 mosaics may have been more phenologically consistent, thus explaining the stronger model fit. Furthermore, the differences in the mosaicking algorithm may have affected the different data qualities. While out of the scope of this study, further research should assess how the different mosaicking methods used for different types of analysis ready data such as the UMD ARD or the Copernicus Sentinel-2 global mosaic (Copernicus 2020) affect the model fit and thus the precision of the ΔAGB estimates.

Even though Landsat based estimates were less precise than Sentinel-2, it was encouraging to see that when using globally available ARD products (e.g., UMD ARD data), the decrease in precision compared to a custom dataset (i.e., Sentinel-2) was only marginal. The main advantage of using Landsat data in combination with NFI data is that it allows us to retroactively estimate AGB dynamics for past NFI inventory cycles, which is not possible for Sentinel-2. Thus, if Sentinel-2 may become the standard for future ΔAGB estimation (GFOI 2020), Landsat data can form the backbone for a retroactive estimation of AGB dynamics. Further efforts should be devoted to exploring synergies between Sentinel-2 and Landsat data concerning the generation of cloud-free mosaics drawing both from Sentinel-2 and Landsat (Shao et al. 2019). Such data products could benefit from the availability of a denser image time series because of the increased chances of cloud-free imagery, thus ensuring a continuous monitoring of forest AGB dynamics.

*3.3 Comparison between using bi- or uni-temporal satellite data*

We found that there was only a small increase in precision when using bi-temporal data compared to uni-temporal data, indicating that ΔAGB could also be estimated when remotely sensed data are available only from the end of the monitoring period. A possible explanation behind such findings is that the magnitude of change in AGB is partly explained by the reflectance properties of the forest canopy in the different developmental stages (AGB gain) or of harvested sites (AGB loss). The quantification of the AGB gains reflect differences in the reflectance due to the phenology of the forest during different developmental stages and thus with different growth patterns. The possibility to estimate AGB losses, on the other hand, might be related to variations in reflectance driven by differences in the cover and floristic compositions of post-harvest ground vegetation. This is in line with the findings by Bergstedt and Milberg (2001) who, using Swedish NFI data, found that post-harvest ground cover vegetation and floristic composition was strongly correlated to the logging intensity (i.e., extracted timber volume). The main advantages of using remotely sensed data for only the end of the monitoring period are: 1) no requirement of availability of two cloud-free mosaics, and thus increases the years for which ΔAGB estimates may be possible, and 2) retroactively estimate ΔAGB for inventory cycles prior to the existence of a certain data source. The latter is particularly relevant for Sentinel-2 data, which are available only since 2015, and thus only changes after 2019 can be estimated based on bi-temporal data. On the other hand, using uni-temporal Sentinel-2 data allows us to estimate ΔAGB that occurred before 2015 (e.g., NFI cycle between 2010 and 2015).

*3.4 Scalability*

When considering the spatial scalability of the methods described in this study to broader scales, one must be aware of a potentially larger heterogeneity in the quality of the image mosaics due to increased cloud cover, seasonal variations, and variations in atmospheric conditions. Such increased variability in the predictor variable is likely to have a detrimental effect on the model fit and thus on the precision of the ΔAGB estimates.

A further aspect to consider is that we used an existing nationwide forest mask to delineate the studied population (i.e., forests) and thus exclude model predictions from non-forest areas. When wanting to extend our method to areas outside of Norway, there is a need for a global forest mask with consistent accuracy. Several options exist nowadays as a global source forest non-forest maps such as the tree cover dataset for the year 2000 produced using Landsat 7 ETM (Hansen et al. 2013), the forest/non-forest map produced from TanDEM-X data (Martone et al. 2018) or the future availability of the WorldCover map at 10 m resolution (ESA 2020). While several masks may be used, further studies should investigate the effect of using different forest masks on the precision of the estimates.

The temporal scalability of the proposed method also is of interest. While we only looked at a single five-year period, the described method can be applied retroactively to improve the precision of time series of $\Delta AGB$ estimates from past inventories, thus allowing us to define more precise baselines. In such a context, Landsat data can prove invaluable as they provide a multi-decadal time-series dating back to 1970s. With that in mind, it remains important to acknowledge the varying data quality for the different Landsat mission.

**4 Conclusion**

This study is the first example of a direct estimation following Tier 3 requirements for ΔAGB using NFI data together with freely available satellite optical data. Based on the results of this study, our conclusion is threefold:

- Satellite optical data can boost the precision of NFI field-based estimates of forest ΔAGB. Such a result is encouraging for future use alongside NFI programs to provide a more precise understanding of the role of forest in the carbon cycle.
- Multi-temporal Sentinel-2 data yielded the most precise results, followed by the use of uni-temporal Sentinel-2.
- Even though Landsat data resulted in a smaller precision of the estimates compared to Sentinel-2, its use for improving past AGB estimates could prove to be very useful.

Future studies should investigate the scalability of the proposed method through space and time.

**Acknowledgements**
We are grateful to Peter Potapov and Matthew Hansen for providing the GLAD Landsat ARD data, to the Copernicus and Landsat programs for providing free access to their data. This study was supported by the Norwegian Institute for Bioeconomy Research (NIBIO) and ERA-GAS INVENT (NRC number 276398).
**References**

Ahlstrøm, A., Bjørkelo, K., & Frydenlund, J. (2014). AR5 klassifikasjonssystem – klassifikasjon av arealressurser [AR5 classification scheme – classification of areal resources] *Rapport fra Skog og landskap [Norwegian]* Available at: https://www.nibio.no/tema/jord/arealressurser/arealressurskart-ar5/klassifikasjonssystem-ar5?locationfilter=true

Astrup, R., Rahlf, J., Bjørkelo, K., Debella-Gilo, M., Gjertsen, A.-K., & Breidenbach, J. (2019). Forest information at multiple scales: development, evaluation and application of the Norwegian forest resources map SR16. *Scandinavian Journal of Forest Research, 34*, 484-496

Baccini, A., Goetz, S.J., Walker, W.S., Laporte, N.T., Sun, M., Sulla-Menashe, D., Hackler, J., Beck, P.S.A., Dubayah, R., Friedl, M.A., Samanta, S., & Houghton, R.A. (2012). Estimated carbon dioxide emissions from tropical deforestation improved by carbon-density maps. *Nature Climate Change, 2*, 182-185

Bergstedt, J., & Milberg, P. (2001). The impact of logging intensity on field-layer vegetation in Swedish boreal forests. *Forest Ecology and Management, 154*, 105-115

Bollandsås, O.M., Ene, L.T., Gobakken, T., & Næsset, E. (2018). Estimation of biomass change in montane forests in Norway along a 1200 km latitudinal gradient using airborne laser scanning: a comparison of direct and indirect prediction of change under a model-based inferential approach. *Scandinavian Journal of Forest Research, 33*, 155-165

Bollandsås, O.M., Gregoire, T.G., Næsset, E., & Øyen, B.-H. (2013). Detection of biomass change in a Norwegian mountain forest area using small footprint airborne laser scanner data. *Statistical Methods & Applications, 22*, 113-129

Breidenbach, J., Granhus, A., Hylen, G., Eriksen, R., & Astrup, R. (2020a). A century of National Forest Inventory in Norway – informing past, present, and future decisions. *Forest Ecosystems, 7*, 46

Breidenbach, J., Waser, L.T., Debella-Gilo, M., Schumacher, J., Rahlf, J., Hauglin, M., Puliti, S., & Astrup, R. (2020b). National mapping and estimation of forest area by dominant tree species using Sentinel-2 data. *Canadian Journal of Forest Research*

Cohen, W.B., Yang, Z., & Kennedy, R. (2010). Detecting trends in forest disturbance and recovery using yearly Landsat time series: 2. TimeSync — Tools for calibration and validation. *Remote Sensing of Environment, 114*, 2911-2924

Copernicus (2020). S2GM User Manual Available at: https://usermanual.readthedocs.io/en/stable/index.html

Csillik, O., Kumar, P., Mascaro, J., O'Shea, T., & Asner, G.P. (2019). Monitoring tropical forest carbon stocks and emissions using Planet satellite data. *Scientific Reports, 9*, 17831